\documentclass{article}
\usepackage{emulateapj}

\slugcomment{To Appear in ApJL}

\begin{document}
\title{A Reappraisal of the Solar Photospheric C/O Ratio}

\author{Carlos Allende Prieto and David L. Lambert}
\affil{McDonald Observatory and Department of Astronomy \\ University of Texas
  \\ RLM 15.308, Austin, TX 78712-1083, \\ USA}
  

\author{Martin Asplund}
\affil{Research School of Astronomy and Astrophysics, Mt. Stromlo Observatory \\
Cotter Road, Weston, ACT 2611, \\ Australia}

\begin{abstract}

Accurate determination of  photospheric solar abundances requires detailed
modeling of the solar granulation and accounting for departures from local
thermodynamical equilibrium (LTE). 
We argue that
the forbidden  C\,{\sc I} line at 8727 \AA\ is largely immune to departures
from LTE, and can be realistically modeled using LTE radiative transfer in
a time-dependent three-dimensional simulation of solar surface
convection. We analyze the [C\,{\sc I}] line in the solar
flux spectrum to derive the abundance
 $\log \epsilon({\rm C})= 8.39 \pm 0.04$ dex. Combining this result
with our parallel analysis of the [O\,{\sc I}] 6300 \AA\ line,
we find  C/O$=0.50 \pm 0.07$, in  agreement with the ratios measured
in the solar corona from gamma-ray spectroscopy and solar energetic particles.

 {\it Subject headings:   convection ---   line: formation
 --- Sun: abundances --- Sun: photosphere}

\end{abstract}

\section{Introduction}

Pursuit of quantitative stellar spectroscopy not infrequently
introduces new puzzles as old puzzles are probed. Recent determinations
of the solar photospheric abundances of carbon and oxygen exemplify this
claim. In a thorough review of light element photospheric abundances,
Holweger (2001) recommends the abundances 
$\log\epsilon$(C) = 8.592 $\pm 0.108$,
and $\log\epsilon$(O) = 8.736 $\pm$ 0.078 on the usual scale with
$\log\epsilon$(H) = 12.0. These results based on 1D 
NLTE analyses of permitted lines of C\,{\sc I} and O\,{\sc I} include
corrections for the effects of solar granulation based on 
2D modeling\footnote{We note that Holweger defines
the granulation corrections as the difference between the 2D result and the
case for the 1D spatial average of the 2D model atmosphere. Holweger's approach
therefore only accounts for the temperature inhomogeneities but not the 
different overall
temperature structures in hydrodynamical and hydrostatic models. Finally,
 we note that
there are systematic differences between 2D and 3D (Asplund et al. 2000b)}.
 Prior to Holweger's reassessment, the accepted carbon
abundance was lower  ($\log\epsilon$(C) = 8.52 $\pm$ 0.06) and that of
oxygen higher ($\log\epsilon$(O) = 8.83 $\pm$ 0.06; Grevesse \& Sauval 1998).
With the reassessment, the carbon-to-oxygen ratio is increased. 
More intriguingly, the errors assigned to the new solar 
abundances would formally allow a carbon-to-oxygen ratio greater than one. 
Such a ratio is exceedingly puzzling in that it implies remarkable abundance 
differences between the photosphere and the corona, the outer planets, 
and unevolved stars  and interstellar gas in the solar neighborhood.
Lowering the 
oxygen abundance and keeping
the high value for the carbon abundance leads also to a conflict with the
strength of observed water and methane bands in the spectra of very cool
stars (Tsuji 2002). Recognizing the controlling influence of the CO
molecule on the partial pressures of carbon and oxygen, the
 simple observation of very weak C$_2$ bands 
and strong TiO bands in sunspot spectra shows that the  carbon-to-oxygen ratio
must be less than unity. Yet, a precise determination of the photospheric
ratio is of great interest.

In this paper, we reconsider the carbon abundance provided by the forbidden
carbon line at 8727 \AA.  Our reanalysis follows our
successful study of the forbidden oxygen line at
6300 \AA, which
gave the abundance $\log\epsilon$(O) = 8.69 $\pm$ 0.05 with a
3D hydrodynamical model photosphere simulating the 
solar granulation (Allende Prieto, Lambert, \& Asplund 2001a).
 The principal advantages of a forbidden line over  permitted
lines are twofold: (i) the line's $gf$-value is most probably more
accurately known than the $gf$-values of the permitted lines, and (ii)
the line is formed very close to LTE, which is not true for the
permitted lines.
We argue that the possible downsides with the [C\,{\sc I}] line, namely its
relative weakness and possible blends, can be adequately addressed.

\begin{figure*}
\begin{center}
\includegraphics[width=10cm,angle=0]{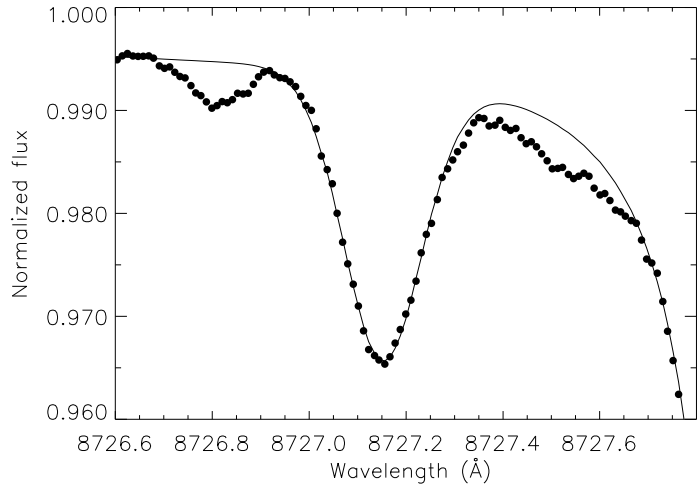}
\figcaption{
Comparison between the observed (filled circles) and synthetic line profiles
for the [C\,{\sc I}] at 8727.1 \AA. Four parameters have been adjusted
to minimize the $\chi^2$: the wavelength of the [C\,{\sc I}] line,
a multiplicative correction to the continuum level,
the strength of the strong Si line at 8728.0 \AA\
($\log [gf \epsilon({\rm Si})]$), and the carbon abundance, which is found to
be $\log \epsilon({\rm C})=8.39 \pm 0.04$ dex.
\label{shifts}}
\end{center}
\end{figure*}

\section{Model Atmospheres and Line Synthesis}

Our line profile calculations follow very closely those of
 Allende Prieto et al. (2001a) for the [O\,{\sc I}] line at 6300 \AA.
We used a three-dimensional time-dependent hydrodynamical simulation of the
solar surface as a model atmosphere (Asplund et al. 2000a and references
therein). The calculated flux profiles for the forbidden [C\,{\sc I}] line
at 8727.1 \AA\ and a Si\,{\sc I} line at 8728 \AA\ result from the average
of a time sequence of flux profiles computed from 100 snapshots, which are
 equally spaced over 50 minutes of solar time.
For each snapshot, the integration over the solar disk makes use
of intensity profiles for  $4 \times 4$  angles, accounting for the
(solid body $v \sin i = 1.9$ km s$^{-1}$)
rotational broadening (Dravins \& Nordlund 1990). The Uppsala opacity
package (Gustafsson et al. 1975 with subsequent updates) was the source of the
continuum opacities, partition functions, ionization potentials, and other
basic data for the line synthesis as well as for the simulation.
Collisional
broadening was evaluated using the Uns\"old's approximation (Uns\"old 1955)
for the [C\,{\sc I}] line and the neighboring Si\,{\sc I} line.

The [C\,{\sc I}] 8727.1 \AA\ line was identified in the solar spectrum
by Lambert \& Swings (1967a). They remarked upon a potential
blending line of Fe\,{\sc I}, 
but deemed its contribution  to be negligible.
Kurucz (1993) calculated $\log gf = -3.93$ for this iron line, which leads
to an equivalent width of $\sim 0.5$ m\AA. Kurucz also
calculated the transition
probabilities for the other 5 lines in the multiplet, and four of them appear
  in the solar spectrum.
 Using the measured equivalent widths ($12 \le W_{\lambda} \le 40$ m\AA)
 to scale the $\log gf$'s of these four lines, we estimate an
 equivalent width of $\sim 0.18 \pm 0.13$ m\AA\footnote{The quoted
 uncertainties in the paper represent 1$\sigma$ values.}.
Nave et al. (1994) in a thorough reinvestigation of the laboratory
Fe\,{\sc I} spectrum did not list the Fe\,{\sc I} line with its
predicted wavelength of 8727.130 \AA, but four stronger lines of
the same multiplet were measured. Three of the four measured lines
 appear relatively unblended in the solar spectrum. Normalizing
  the line strengths measured by Nave et al.
to the  relative LS-coupling strengths, we can scale the
observed equivalent widths and obtain a third estimate of the
equivalent width of  the 8727.130 \AA\ line: $0.11 \pm 0.02$ m\AA.
Given that the solar feature
has an equivalent width of about 6 m\AA, the Fe\,{\sc I} contribution
is very small and can be neglected.
Gustafsson et al. (1999) eliminate the CN Red
system as another contributor of a blend.

\begin{figure*}
\begin{center}
\includegraphics[width=7cm,angle=0]{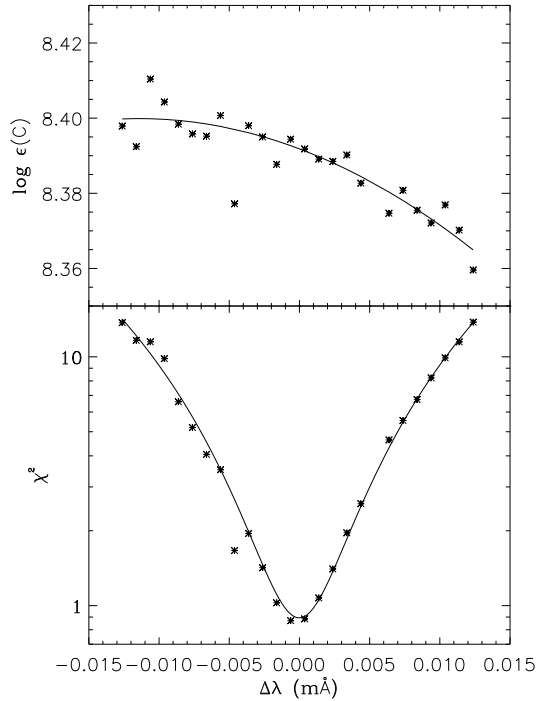}
\figcaption{
Derived carbon abundance and minimum value of the  reduced $\chi^2$ as a function
of the displacement of the central wavelength for the [C\,{\sc I}] transition.
The best fit is achieved for
$\lambda_{\rm [C\,{\sc I}]}= 8727.139 \pm 0.004$ \AA, which is adopted as the
 zero wavelength shift in the Figure. The solid lines are least-squares 
 second-order polynomials fit to the data.
\label{shifts}}
\end{center}
\end{figure*}

The predicted [C\,{\sc I}] rest wavelength is 8727.126 \AA\ with
an uncertainty of about 0.015 \AA\ (Moore 1993).
After correction for 
the gravitational redshift, the center of the
 feature  in the solar spectrum is observed
at $8727.133 \pm 0.009$ \AA\
(as determined with the method described in Allende Prieto et al. 2002).
For weak lines, we expect  a significant
convective blueshift (Allende Prieto \& Garc\'{\i}a L\'opez 1998).
Therefore, the predicted wavelength is likely to be in error by as much
as its full quoted uncertainty.
Different calculations
report results for the Einstein $A$-value which have converged to
$A$ = 0.640 s$^{-1}$ or $\log gf$ = $-$8.136  -- 
see  Galavis et al. (1997), and  Hibbert et al. (1993).
A strong Si\,{\sc I} line affects the continuum
in the vicinity of the  [C\,{\sc I}] line. The NIST database quotes  a
measured
wavelength of 8728.011 \AA\ for this transition.
We adopted $\log gf=-0.37$ in order to match the observed line
reasonably well.

\section{Comparison with the Observed Spectrum}

We compare our calculated line profiles with the Fourier transform solar flux
spectrum of Kurucz et al. (1984).
The [C\,{\sc I}] lies in the
 scan \# 13, which includes the telluric O$_{2}$ line at 6883.8335 \AA,
 used to normalize the frequency scale.
 This scan  has a spectral resolving
 power in excess of half a million, and a  maximum signal-to-noise of 2600.

The Si\,{\sc I} line dominates the shape of the local
continuum in the region
of the [C\,{\sc I}] line. The  abundance of silicon has been
recently studied using  the same 3D model atmosphere employed here
(Asplund 2000). Recent 1D
NLTE calculations suggest that photospheric Si\,{\sc I} lines are not much
affected by departures from LTE (Wedemeyer 2001).
Since the line's $gf$-value is uncertain, we considered the factor
 $\log [gf \epsilon({\rm Si})]$ and a multiplicative
 correction to the local continuum level as
 free parameters with which to adjust the continuum in the
 vicinity of the [C\,{\sc I}] line.
A variation in
the central wavelength of the Si\,{\sc I} line has a similar effect as a
change in the strength of the line.

  The velocity shift between
 Kitt Peak National Observatory
 and the Sun at the time of the observations has been corrected typically
 within several m s$^{-1}$, but other factors (setting of frequency scale in
 FTS, conversion to air wavelengths, correction of individual scans with
 partial wavelength coverage, change of velocity shift over long
 integrations, etc.) can degrade the absolute accuracy of the wavelength scale
 up to 0.1 km s$^{-1}$ (Kurucz et al. 1984).
 The  gravitational redshift for photospheric photons intercepted at Earth
  is $V_g \simeq G M_{\odot}/(R_{\odot} c) \simeq 0.6336$ km s$^{-1}$.
 Seasonal variations produced by the ellipticity of Earth's orbit
 are about 0.1 m s$^{-1}$, Earth's surface gravity induces a
 correction of about 0.2 m s$^{-1}$,  and uncertainties  in
 the solar radius and $GM_{\odot}$ can hamper $V_g$ by about 0.3
 m s$^{-1}$. We are using
 a hydrodynamical model atmosphere which has been shown to predict the
 convective shifts of photospheric lines of weak and moderate strength within
 0.1 km s$^{-1}$ (Asplund et al. 2000a). Therefore, we should be able
 to predict the measured
 wavelength of the  [C\,{\sc I}] with an uncertainty of about 0.15 km s$^{-1}$
 ($\simeq 0.004$ \AA), an estimate significantly
 smaller than for Moore's wavelength. We consequently adjust the wavelength of 
 the line to best match the observations.

  We   proceed to compare our
 calculated profiles to the solar observations adopting different values for
 the wavelength of the  forbidden line $\lambda_{\rm [C\,{\sc I}]}$.
 For each adopted $\lambda_{\rm [C\,{\sc I}]}$, we minimize the
 $\chi^2$ between  the observed and calculated spectra
 by changing the abundance of carbon $\log \epsilon({\rm C})$, a multiplicative
 correction factor to the local continuum,
 and $\log [gf \epsilon({\rm Si})]$.
 There are  obvious blends on both sides
 of the [C\,{\sc I}] line. The  line at 8726.8 \AA\ is a CN line
 (Gustafsson et al. 1999), but other lines are unidentified.
We believe there are no blends within the [C\,{\sc I}] profile,
but, if we are in error, the determined carbon abundance is an upper limit.

First, we fitted 1 \AA\ around the [C\,{\sc I}]
 feature, between 8726.6 and 8727.6 \AA\ in the solar atlas, using the
 Nelder-Mead simplex method (Nelder \& Mead 1965; Press et al. 1986).
 We obtain the best fit, illustrated in Fig. 1, for
 $\log \epsilon({\rm C})= 8.39$ dex.
We emphasize that the fit is achieved without invoking micro- or
macroturbulence; the line profile is predicted from the convective
flows without additional parameters.
 By selecting a narrow interval
excluding the core of the Si\,{\sc I} blending line, we are in
effect adjusting the line's damping constant to get the optimum fit to the
wing around the [C\,{\sc I}] line; the line depth in a damping wing
scales as the product of the abundance and the damping constant.
Next, we repeated  the experiment for a more
 restricted wavelength interval, between 8726.9 and  8727.25 \AA.
The carbon abundance changed by less than 0.01 dex,  the
 multiplicative
factor to adjust the continuum level changed by only 0.0003, and
 the best values for the wavelength of the [C\,{\sc I}] by less than
 0.001 \AA.
 Our best fit yielded a reduced
 $\chi^2$ (37 frequencies and 4 degrees of freedom) of 0.87,
 and, therefore, the chance probability for this is $P=0.3$. These figures are
 based on the signal-to-noise ratio of $\sim 2100$
derived by  examination of
 an apparently clean segment of the
 spectrum around 8731.7 \AA.
 Fig. 2 shows the variation of the reduced $\chi^2$ and the carbon abundance
 as a function of the central wavelength for the
 the [C\,{\sc I}] transition, when the more restricted wavelength interval
 is considered.

 The  rest wavelength of the [C\,{\sc I}] line
 that minimizes the $\chi^2$ is $8727.139 \pm 0.004$ \AA, after correction
for the gravitational redshift. The synthetic spectra predict a blueshift of
about 0.25 km s$^{-1}$ for the line, i.e., a laboratory wavelength of
8727.146 \AA. This differs from Moore's wavelength of 8727.126 \AA\ obtained
not from laboratory measurements of the line but from ultraviolet
lines connecting the ground configuration levels to excited levels. In the
introductory remarks to her table of C\,{\sc I} energy levels,
she draws attention to Kaufman \& Ward's (1966) revisions 
of ultraviolet wavelengths which
imply revised energies for the ground configuration. These
revisions predict a wavelength of 8727.141 \AA\ in good agreement with our
solar-based wavelength, even though the uncertainty remains at about
 0.01 \AA.
 
The internal uncertainty in the fit for the carbon abundance can be estimated
 as 0.01 dex by confronting the expected accuracy in
 $\lambda_{\rm [C\,{\sc I}]}$ (0.004 \AA, which includes the observational
 share of the error), with Fig. 2.
Judging from the scatter in the most recently calculated  
$A-$values,
 the derived carbon abundance is hardly affected by this factor
 ($\sigma[\log \epsilon({\rm C})] \simeq 0.005$ dex).
Other systematic errors may arise from
 the uncertainties in the
 adopted continuum opacities, equation of state, etc.,
 probably about 0.02 dex. Finally, if the contribution
  of the Fe\,{\sc I} line discussed in \S 2  to the observed absorption is
 close to the lower limit of our estimates, the effect on the derived carbon
 abundance is truly negligible, but if it is close to the upper limit, that
 could decrease $\log \epsilon({\rm C})$ by $\simeq 0.03$ dex.
  Conservatively, we derive $\log \epsilon({\rm C}) = 8.39 \pm 0.04$ dex.
As shown by St\"urenburg \& Holweger (1990), 
the forbidden line at 8727.1 \AA\ is immune to departures from LTE, and 
therefore our LTE calculation of the line formation in a 3D LTE model 
is likely to provide a reliable abundance.

It is interesting to compare our 3D result to
a standard  analysis using 
one-dimensional model atmospheres. Use of the solar
semi-empirical model atmosphere
derived by Allende Prieto et al. (2001b), and the micro- and
macro-turbulence derived with the model ($\xi=1.1$ km s$^{-1}$, macro=1.54
km s$^{-1}$),   leads to
$\log \epsilon({\rm C}) = 8.47$ dex, the
Holweger-M\"uller model gives $\log \epsilon({\rm C}) =  8.48$ dex,
and a flux-constant MARCS model atmosphere $\log \epsilon({\rm C}) =  8.41$ dex.

\section{The Solar Carbon Abundance and C/O Ratio}

Our determination of the solar carbon abundance 
($\log \epsilon({\rm C}) = 8.39$) from the forbidden line at
8727.1 \AA\ is lower than 
the most recent estimates based on this line. 
Our lower abundance is mainly caused for two reasons: a revised
 $f-$value ($\simeq -0.07$ dex), and the use of a 3D model atmosphere
($\simeq -0.08$ dex).

Other indicators of the solar photospheric carbon abundance are available. 
 C\,{\sc I},  CH, and C$_2$ lines
have received significant attention in the literature and suggest, when 
analyzed with empirical or
theoretical 1D model atmospheres, higher abundances 
than our determination from
the [C {\sc I}] line  (see, e.g. Grevesse et al. 1991). Pending a 
full reanalysis with the
3D model, we note that the C abundance from the molecular lines
will be reduced in the change from a 1D to 3D model due to the high
temperature sensitivity of molecule formation and the presence of
cooler gas in the 3D model (Asplund \& Garc\'{\i}a P\'{e}rez 2001). 
To assess the effect on the
C\,{\sc I} lines, we selected the weakest and least blended 13 lines
in overlapping lists of solar lines used previously by Bi\'emont et al.
(1993) and St\"urenburg \& Holweger (1990, also Holweger 2001).
Analysis of the published equivalent widths for the center of the solar
disk spectrum with the $gf$-values in the NIST database  and the 3D model gives
the LTE abundance $\log\epsilon$(C) $= 8.44 \pm 0.06$. NLTE 3D calculations are
not available, but if the
1D corrections for NLTE effects from St\"urenburg \&
Holweger are adopted, the abundance is lowered to 
$\log\epsilon$(C) = 8.42 $\pm$ 0.06,
a value in harmony with our result from the [C\,{\sc I}] line. 
 Our preliminary conclusion is that the lower carbon
abundance  is not contradicted by other indicators.

Combining the  updated solar abundances of carbon  and
 oxygen   from the forbidden lines, we obtain
a number-density ratio C/O$=0.50 \pm 0.07$.  
Independent estimates of the C/O ratio
can be obtained for the solar wind and the corona. The C/O ratio
has been  measured from
 gamma-rays produced by the interaction of particles produced
in  solar flares with the surrounding medium. 
Solar Maximum Mission (SMM) data for a flare observed in
1981 yielded  C/O $=0.42 \pm 0.09$ (Murphy et al. 1990; Fludra et al. 1998). 
Ramaty et al. (1996)
derived  $0.35 <$ C/O $< 0.44$ for 19 events observed by SMM.
More recently Murphy et al. (1997) obtained $0.54 \pm 0.04$ from CGRO data
of a flare observed in 1991. Solar energetic particles   
detected near 
Earth can themselves be studied to determine relative abundances.
Reames (1995) reviews coronal abundances derived from solar
energetic particles. Compared to photospheric abundances, 
elements with a low first-ionization potential (FIP)
($\lesssim 9$ eV) are enhanced in the corona, 
but those with a higher FIP are not.
As carbon and oxygen have a similar, high, FIP, this effect is not
expected to disturb their ratio greatly from its photospheric value. 
For `gradual' events associated with coronal mass ejections, the C/O ratio
shows very little variation from one event to another: Reames gives C/O =
0.465 $\pm$ 0.009. For events associated with flares, C/O = 0.434 $\pm$ 0.030.
These results are in good agreement with our `forbidden' photospheric value.

The revised abundances of carbon and oxygen in the Sun compare very well 
with the  abundances (gas plus dust) derived from recombination lines in 
the Orion nebula (Esteban et al. 1998): 
$\log \epsilon ({\rm C})= 8.49 \pm 0.12$, and
 $\log \epsilon ({\rm O})= 8.72 \pm 0.07$ (C/O $= 0.59$).
 They are also in good agreement
with the average values  found in B-type stars in the field:
$\log \epsilon ({\rm C})= 8.31$, and $\log \epsilon ({\rm O})= 8.58$ 
(C/O $=0.54$), as well as in clusters:
 $\log \epsilon ({\rm C})= 8.35$, and $\log \epsilon ({\rm O})= 8.69$  
(C/O $=0.46$)
(Kilian 1992, 1994; Adelman Robinson \& Wahlgren 1993; Gies \& Lambert 1992;
compiled  by Snow \& Witt 1996). 

Revision of the solar C and O abundances using their
forbidden lines brings these abundances into line with
recent results for B stars and the Orion nebula. One may
speculate that this close correspondence in compositions
may extend to other  elements, for example, nitrogen, 
through revisions to
either the solar, stellar or nebular abundances.
Unfortunately, the [N\,{\sc I}] lines are too
weak for positive identification in the solar spectrum
(Lambert \& Swings 1967b). Nitrogen is represented by
weak N\,{\sc I} lines for which Holweger (2001) gives
$\log\epsilon$(N) = 7.93 $\pm$ 0.11 after including NLTE effects
and a correction for granulation. Combining the N/O ratios measured
for solar energetic particles with the photospheric  abundance of oxygen
gives a nitrogen abundance which is about 0.1 dex lower than Holweger's, 
and in better agreement with nearby B stars and the gas in Orion.

\acknowledgements
We thank Toby Owen for provoking us to  extend our analyses
of solar forbidden lines.
NSO/Kitt Peak FTS data used here were produced by NSF/NOAO. 
This work has been partially funded by the 
US National Science Foundation (grant AST 00-86321),
the Robert A. Welch Foundation of Houston, Texas, 
the Swedish Natural Science
Foundation (grant NFR F990/1999), and the Royal Swedish Academy
of Sciences.


\begin{thebibliography}{}

\bibitem[Adelman, Robinson, \& Wahlgren(1993)]{1993PASP..105..327A} 
Adelman, S.~J., Robinson, R.~D., \& Wahlgren, G.~M.\ 1993, \pasp, 105, 327. 



\bibitem[Allende Prieto \& Garcia Lopez(1998)]{1998A&AS..129...41A} Allende 
Prieto, C., \& Garc\'{\i}a L\'opez, R.~J.\ 1998, \aaps, 129, 41. 

\bibitem[Allende Prieto, Lambert, Tull, \& 
MacQueen(2002)]{2002ApJ...566L..93A} Allende Prieto, C., Lambert, D.~L., 
Tull, R.~G., \& MacQueen, P.~J.\ 2002, \apjl, 566, L93. 

\bibitem[Allende Prieto, Lambert, \& Asplund(2001)]{2001ApJ...556L..63A} 
Allende Prieto, C., Lambert, D.~L., \& Asplund, M.\ 2001a, \apjl, 556, L63. 

\bibitem[Allende Prieto, Barklem, Asplund, \& Ruiz 
Cobo(2001)]{2001ApJ...558..830A} Allende Prieto, C., Barklem, P.~S., 
Asplund, M., \& Ruiz Cobo, B.\ 2001b, \apj, 558, 830. 

\bibitem[2001]{asplund01}
Asplund, M., \& Garc\'{\i}a P\'erez, A.E. 2001, \aap, 372, 601.

\bibitem[Asplund(2000)]{2000A&A...359..755A} Asplund, M.\ 2000, \aap, 359, 
755. 

\bibitem[Asplund et al.(2000)]{2000A&A...359..729A} Asplund, M., Nordlund, 
{\AA}., Trampedach, R., Allende Prieto, C., \& Stein, R.~F.\ 2000a, \aap, 
359, 729. 

\bibitem[Asplund, Ludwig, Nordlund, \& Stein(2000)]{2000A&A...359..669A} 
Asplund, M., Ludwig, H.-G., Nordlund, {\AA}., \& Stein, R.~F.\ 2000b, \aap, 
359, 669. 

\bibitem[Biemont, Hibbert, Godefroid, \& Vaeck(1993)]{1993ApJ...412..431B} 
Bi\'emont, E., Hibbert, A., Godefroid, M., \& Vaeck, N.\ 1993, \apj, 412, 431.





\bibitem[]{}  
Dravins, D., \& Nordlund, \AA. 1990, \aap, 228, 203. 

\bibitem[Esteban, Peimbert, Torres-Peimbert, \& 
Escalante(1998)]{1998MNRAS.295..401E} Esteban, C., Peimbert, M., 
Torres-Peimbert, S., \& Escalante, V.\ 1998, \mnras, 295, 401. 

\bibitem[Fludra et al.(1999)]{1999mfs..conf...89F} Fludra, A.~et al.\ 1999, 
The many faces of the sun: a summary of the results from NASA's Solar 
Maximum Mission., 89. 

\bibitem[Galavis, Mendoza, \& Zeippen(1997)]{1997A&AS..123..159G} Galavis, 
M.~E., Mendoza, C., \& Zeippen, C.~J.\ 1997, \aaps, 123, 159. 

\bibitem[Gies \& Lambert(1992)]{1992ApJ...387..673G} Gies, D.~R.,~\& 
Lambert, D.~L.\ 1992, \apj, 387, 673. 



\bibitem[Grevesse et al.(1991)]{1991A&A...242..488G} Grevesse, N., Lambert, 
D.~L., Sauval, A.~J., van Dishoek, E.~F., Farmer, C.~B., \& Norton, R.~H.\ 
1991, \aap, 242, 488. 


\bibitem[Grevesse \& Sauval(1998)]{1998SSRv...85..161G} Grevesse, N.~\& 
Sauval, A.~J.\ 1998, Space Science Reviews, 85, 161. 

\bibitem{} Gustafsson B., Bell R.A., Eriksson K., \& Nordlund \AA ., 1975,
\apj, 42, 407.

\bibitem[Gustafsson et al.(1999)]{1999A&A...342..426G} Gustafsson, B., 
Karlsson, T., Olsson, E., Edvardsson, B., \& Ryde, N.\ 1999, \aap, 342, 426. 



\bibitem[Hibbert, Biemont, Godefroid, \& Vaeck(1993)]{1993A&AS...99..179H} 
Hibbert, A., Bi\'emont, E., Godefroid, M., \& Vaeck, N.\ 1993, \aaps, 99, 179. 


\bibitem[Holweger(2001)]{2001sgc..conf...23H} Holweger, H.\ 2001, Joint 
SOHO/ACE workshop Solar and Galactic Composition, AIP Conf. Proc. 598, ed.
R. F. Wimmer-Schweingruber, p. 23. 

\bibitem[]{} Kaufman, V., \& Ward, J. F. 1966, JOSA, 56,  1591.

\bibitem[Kilian(1994)]{1994A&A...282..867K} Kilian, J.\ 1994, \aap, 282, 
867. 

\bibitem[Kilian(1992)]{1992A&A...262..171K} Kilian, J.\ 1992, \aap, 262, 
171. 




\bibitem[Kurucz(1993)]{1993KurCD..18.....K} Kurucz, R.\ 1993, SYNTHE 
Spectrum Synthesis Programs and Line Data.~Kurucz CD-ROM No.~18.~Cambridge, 
Mass.: Smithsonian Astrophysical Observatory, 1993., 18. See 
{\tt http://kurucz.harvard.edu/} 

\bibitem[Kurucz, Furenlid, \& Brault(1984)]{1984sfat.book.....K} Kurucz, 
R.~L., Furenlid, I., Brault, J., \& Testerman, L.\ 1984, National Solar Observatory Atlas, 
(Sunspot, New Mexico: National Solar Observatory). 


\bibitem[Lambert \& Swings(1967)]{1967SoPh....2...34L} Lambert, D.~L.,~\& 
Swings, J.~P.\ 1967a, \solphys, 2, 34. 

\bibitem[Lambert \& Swings(1967)]{1967Obs....87..113L} Lambert, D.~L.,~\& 
Swings, J.~P.\ 1967b, The Observatory, 87, 113 



\bibitem[]{} Moore, C. E. 1993, Tables of Spectra of H, C, N, and O Atoms
and Ions (Boca Rat\'on: CRC Press).

\bibitem[]{} Murphy, R. J., Share,G. H., Grove, J. E., Johnson, W. N., Kinzer, 
R. L., Kurfess, J. D., Strickman, M. S., \& Jung, G. V. 1997, \apj, 490, 883.

\bibitem[Murphy, Share, Letaw, \& Forrest(1990)]{1990ApJ...358..298M} 
Murphy, R.~J., Share, G.~H., Letaw, J.~R., \& Forrest, D.~J.\ 1990, \apj, 
358, 298. 


\bibitem[Nave et al.(1994)]{1994ApJS...94..221N} Nave, G., Johansson, S., 
Learner, R.~C.~M., Thorne, A.~P., \& Brault, J.~W.\ 1994, \apjs, 94, 221. 

\bibitem[]{} Nelder, J. A., \& Mead, R., 1965, Comput. J., 7, 308.


\bibitem[]{} Press, W. H., Flannery, B. P., Teukolsky, S. A., 
\& Vetterling, W. T. 1986, Numerical Recipes (Cambridge: Cambridge Univ. Press).

\bibitem[]{} Ramaty, R., Mandzhavidze, N., \& Kozlovsky, B 1996, 
in AIP Conf. Proc. 374,
High-Energy Solar Physics, ed. R. Ramaty, N. Mandzhavidze, \& X.-H. Hua 
(New York: AIP), 172.

\bibitem[]{} Reames, D. V. 1995, Adv. Space Res. 15, 41.



\bibitem[Snow \& Witt(1996)]{1996ApJ...468L..65S} Snow, T.~P., \& Witt, 
A.~N.\ 1996, \apjl, 468, L65. 


\bibitem[Stuerenburg \& Holweger(1990)]{1990A&A...237..125S} St\"urenburg, 
S.,~\& Holweger, H.\ 1990, \aap, 237, 125. 


\bibitem[Tsuji(2002)]{2002astro.ph..4401T} Tsuji, T.\ 2002,
preprint
(astro-ph/0204401)

\bibitem[Unsold(1955)]{1955QB461.U55......} Uns\"old, A.\ 1955, Physik 
der Sternatmosph\"aren, (Berlin: Springer). 

\bibitem[Wedemeyer(2001)]{2001A&A...373..998W} Wedemeyer, S.\ 2001, \aap, 
373, 998. 




\end{thebibliography}
\end{document}